\documentclass[conference,a4paper]{IEEEtran}
\usepackage{cite}
\usepackage{amsmath,amssymb,amsfonts}
\usepackage{algorithmic}
\usepackage{graphicx}
\usepackage{textcomp}
\usepackage{xcolor}
\def\BibTeX{{\rm B\kern-.05em{\sc i\kern-.025em b}\kern-.08em
    T\kern-.1667em\lower.7ex\hbox{E}\kern-.125emX}}
\begin{document}

\title{Cascaded Convolutional Neural Networks with Perceptual Loss for Low Dose CT Denoising
}

\author{\IEEEauthorblockN{Sepehr Ataei}
\IEEEauthorblockA{\textit{Electrical and Computer Engineering} \\
\textit{Ryerson University}\\
Toronto, Canada \\
s2ataei@ryerson.ca}
\and
\IEEEauthorblockN{Javad Alirezaie}
\IEEEauthorblockA{\textit{Electrical and Computer Engineering} \\
\textit{Ryerson University}\\
Toronto, Canada \\
javad@ryerson.ca}
\and
\IEEEauthorblockN{Paul Babyn}
\IEEEauthorblockA{\textit{Dept. Of Medical Imaging} \\
\textit{University of Saskatoon}\\
Saskatoon, Canada \\
Paul.babyn@saskatoonhealthregion.ca}

}

\maketitle

\begin{abstract}
Low Dose CT Denoising research aims to reduce the risks of radiation exposure to patients. Recently researchers have used deep learning to denoise low dose CT images with promising results. However, approaches that use mean-squared-error (MSE) tend to over smooth the image resulting in loss of fine structural details in low contrast regions of the image. These regions are often crucial for diagnosis and must be preserved in order for Low dose CT to be used effectively in practice.  In this work we use a cascade of two neural networks, the first of which aims to reconstruct normal dose CT from low dose CT by minimizing perceptual loss, and the second which predicts the difference between the ground truth and prediction from the perceptual loss network. We show that our method outperforms related works and more effectively reconstructs fine structural details in low contrast regions of the image. 
\end{abstract}

\begin{figure*}[htbp]
\centerline{\includegraphics[width=\textwidth]{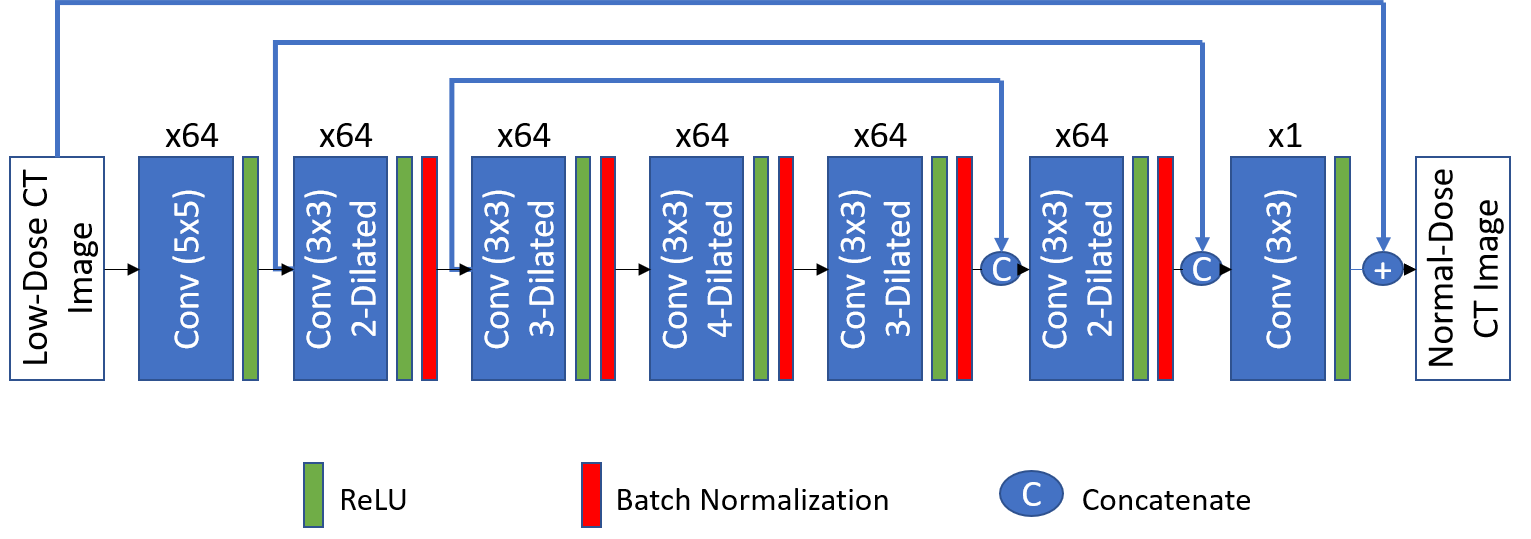}}
\caption{DRL Model Architecture }
\label{model}
\end{figure*}
\section{Introduction}
Radiation exposure is one of the biggest drawback of X-ray computed tomography (CT) \cite{Gonzlez2009ProjectedCR}. Excessive exposure to radiation can have negative side effects and in some cases cause cancer. Unfortunately, when reducing radiation dose the signal to noise ratio (SNR) of CT images reduce considerably, resulting in poor diagnostic accuracy \cite{MARTIN19991}. The research community has proposed various noise reduction techniques as solutions. In the past, researchers were focused on iterative reconstruction (IR) algorithms for Low Dose CT (LDCT) image reconstruction. However, these algorithms were computationally expensive and caused artifacts in CT images and thus their use in practice was not realized \cite{doi:10.1148/rg.344135128}. Further, these methods required raw projection data and as a result were scanner dependent. On the other hand, image post-processing techniques do not require projection data and thus offer a scanner independent and more robust solution.

In the past few years Deep Convolutional Neural Networks have shown their prowess in many computer vision and image processing tasks. In medical imaging, these networks have achieved state-of-the-art in segmentation, classification, denoising and more. Adding more layers to these networks has been shown to improve performance by aiding the optimization of a larger feature space. However, training of deep networks has proven to be problematic due to the vanishing and/or exploding gradient problem \cite{279181}. To solve this, residual networks have been proposed. He et al. showed that simply increasing the number of layers in a network is not an effective way to improve model accuracy. In fact, accuracy will degrade past a certain depth \cite{7780459}. They proposed residual learning, in which the input to a residual block is added to the output of the same block which may contain any number of traditional layers. The output of the residual block is then used as input to the next. This allows information from early layers in the network to more easily propagate to the deeper layers and allow the network to learn low level, medium level and high level features. This advantage is especially useful in image denoising, as preserving full detail of the original noisy image is especially important. 
Generative Adversarial Networks (GANs) were introduced by Goodfellow et al. in 2014. This architecture includes a generator network which learns the desired data distribution G and a discriminator network which estimates the probability that a sample came from the training data rather than G. While training, the networks learn simultaneously as the discriminator improves at distinguishing real samples from fake samples, the generator must then learn to fool the discriminator. This causes the generator to produce better and better samples until the discriminator is no longer able to distinguish between G and training samples (50\% accuracy). However, for complex images GANs have difficulty in training due to the vanishing gradient on the generator \cite{Arjovsky:2017:WGA:3305381.3305404}. Yang et al. used Wasserstein GAN to overcome vanishing gradients and successfully generated normal dose CT (NDCT) from LDCT. They showed that using MSE alone as an objective function over-smooths the output and causes loss of important structural detail. For this purpose, they use a combination of MSE and perceptual loss \cite{8340157}.
In general, denoising algorithms struggle in areas of low contrast where  MSE based methods tend to smooth low contrast structures into the background while GAN based methods tend to remove them as noise \cite{8340157}. To address these issues, Ansari et al. proposed a dilated residual CNN with perceptual loss to effectively propagate low level and high level features through the network and maintain perceptual similarity between predicted and ground truth images \cite{GholizadehAnsari2019DeepLF}. Wu et al. propose a cascaded CNN which simplifies the denoising problem by predicting difference images aiming to improve the output after each cascade level. In this work, we take inspiration from \cite{GholizadehAnsari2019DeepLF} and \cite{Wu2017ACC} to propose a new cascaded network structure (Figure \ref{arch}) which combines MSE based difference image denoising and perceptual loss image reconstruction. We show that the results of the cascaded networks outperform using each network independently, and that our approach is more effective than the original cascade architecture in \cite{Wu2017ACC}. 

 \section{Background}

 \subsection{Perceptual Loss}
 Yang et al. define a perceptual loss based on features extracted by the pre-trained VGG-19 network. They use the output of the 16th convolutional layer as the extracted features, and calculate a mean squared error distance between the features of the ground truth image and corresponding denoised image. In this work we use the VGG-16 network \cite{Simonyan2014VeryDC} to extract features from the last convolutional layer in blocks 1,2,3 and 4 and calculate an averaged mean squared error distance across all feature maps.

 Perceptual loss is effective for the denoising task because it more accurately models the human visual system than MSE. MSE only compares per pixel difference between two images and does not take into account high level features. Deep CNNs such as VGG-16 are able to model the human visual system because they learn features to accurately describe the natural images they are trained on. We take advantage of this artificial visual understanding by penalizing our network when extracted features are dissimilar\cite{inproceedingspercep}.

\section{Methods}

\subsection{Denoising Model}

We can represent an image denoiser $G$ mathematically as a function that maps LDCT to NDCT ,where $\boldsymbol{z} \in \mathbb{R}^{N \times N}$ is a LDCT image and $\boldsymbol{x} \in \mathbb{R}^{N \times N}$ is a NDCT image.

\begin{equation}
G: z \rightarrow x
\end{equation}

Although  noise in raw x-ray measurements can be modelled as a combination of Poisson quantum noise and Gaussian electronic noise, reconstructed CT images do not have a well-defined noise distribution and the noise is non-uniformly distributed across the image. Deep neural networks are advantageous here as they can efficiently learn high-level features and accurate data representations given a sufficiently large training set.

\subsection{Dataset Preparation }
We train and evaluate our methods on the AAPM Low Dose CT Grand Challenge dataset. The dataset contains abdominal CT scans of 10 patients at 1mm slice thickness.  Scans were taken with 120kVP tube voltage  and  an effective dose of 200 mAs .   The  low-dose  images were simulated by adding Poisson noise to the projection data in order to estimate quarter dose acquisition. We extracted 281660 patches of size 64x64 from 8 patients for training, and used the remaining 844 from the remaining two patients for testing. Due to the fully convolutional nature of the proposed networks we are able to train on patches and test on full sized images (512 x 512). This helps to reduce computational load and increase the number of training samples. 

\begin{figure*}[htbp]
\centerline{\includegraphics[width=\textwidth]{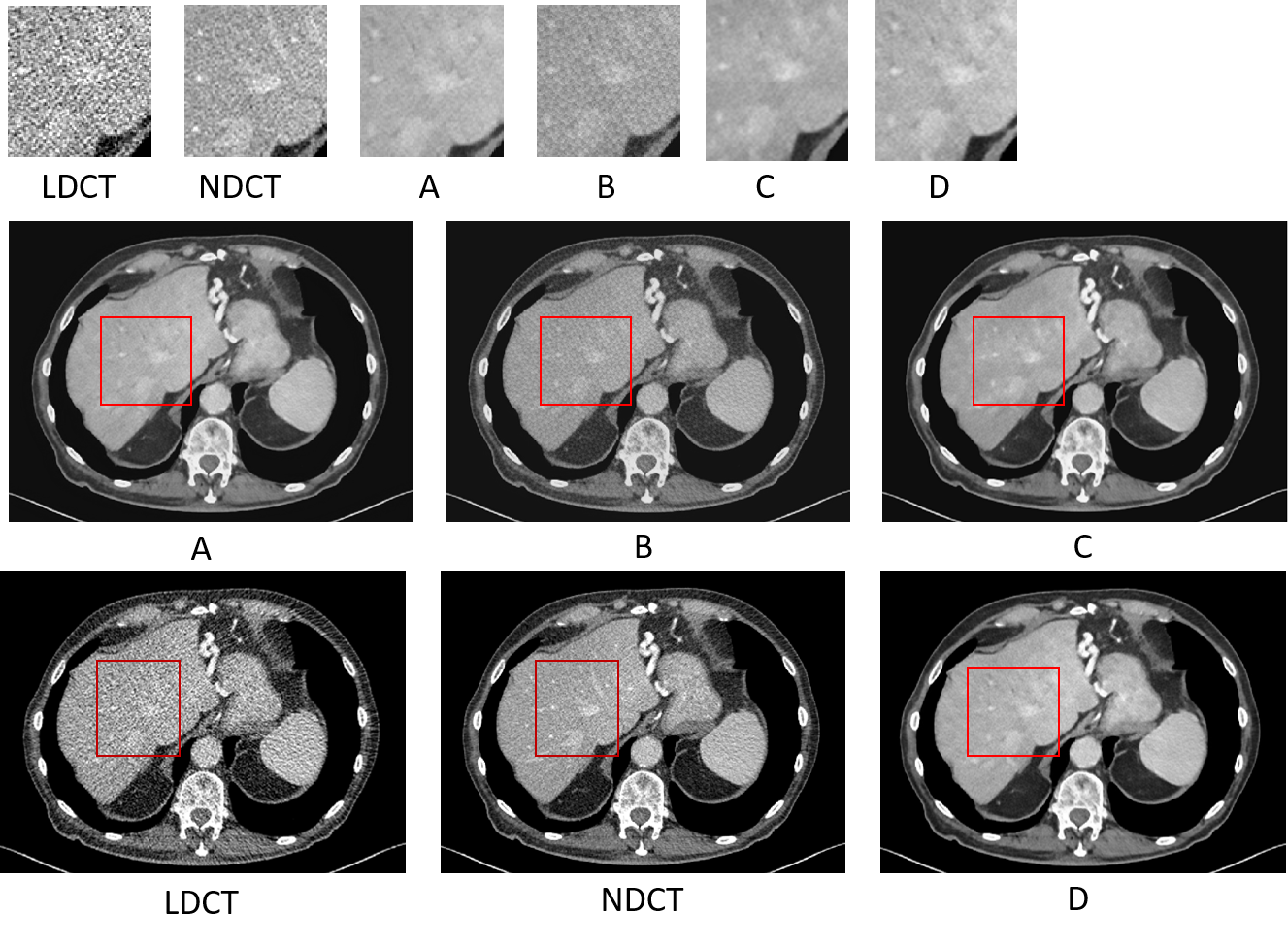}}
\caption{Denoising result on lung test image. A) MSE-based difference image network B) DRL Network with Perceptual Loss, C) Cascade of A and B, LDCT (Low Dose CT), NDCT (Normal Dose CT), D) Two-level cascade of A }
\label{comp}
\end{figure*}

\begin{figure}[htbp]
\centerline{\includegraphics[width=\columnwidth]{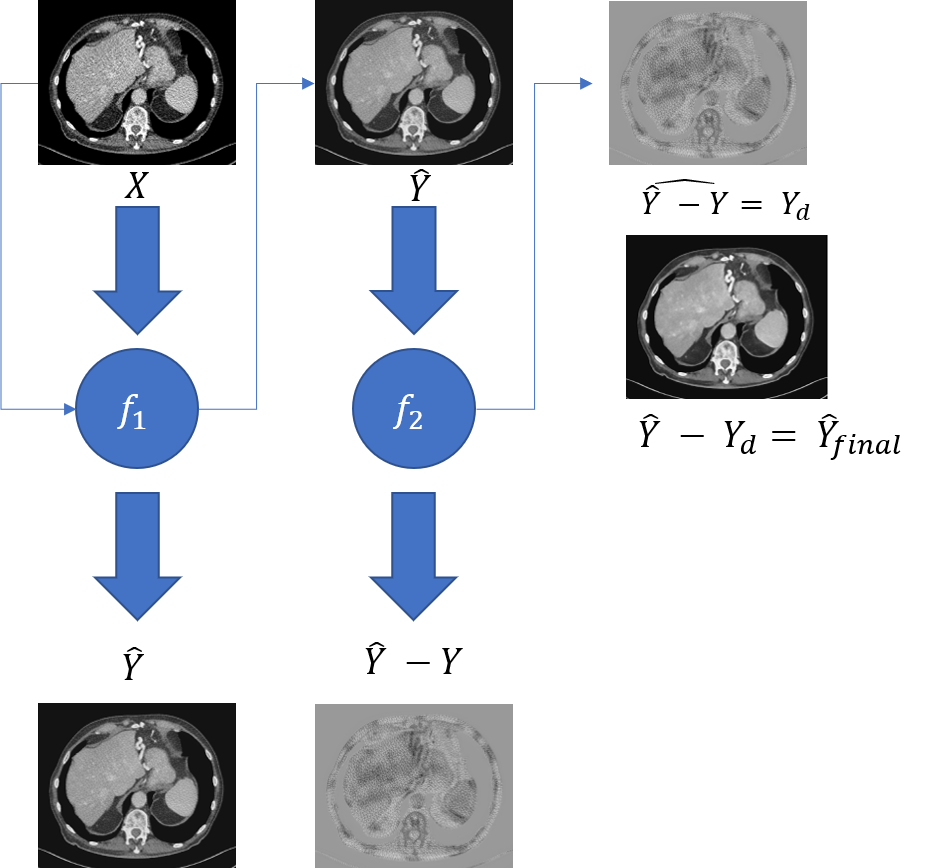}}
\caption{Proposed Denoising System Architecture}
\label{arch}
\end{figure}

\subsection{Objective Functions}
 We take advantage of a combination of MSE and perceptual loss in order to improve PSNR while maintaining structural details and enhancing contrast. For MSE we minimize the loss function $L$ with respect to model parameters $(\theta)$ where $\left\{\left(x_{i}, y_{i}\right)\right\}_{i=1}^{N}$
are NDCT and LDCT image pairs respectively.

\begin{equation}
{L_{MSE}}(\theta)=\frac{1}{N} \sum_{i=1}^{N}\left\|f\left(x_{i} ; \theta\right)-y_{i}\right\|_{F}^{2}
\end{equation}

The perceptual loss $L_{p}(\theta)$ is defined as:

\begin{equation}
{L}_{P}(\theta)=\sum_{i=1}^{4}\frac{1}{h_{i} w_{i} d_{i}}\left\|\phi_{i}(\hat{y}(\theta))-\phi_{i}(y)\right\|^{2}
\end{equation}

where  $\hat{y}(\theta) $ is a denoised image and $y$ is the corresponding ground truth image. We extract feature maps $\phi_{i}$  from block $i$ of the pre-trained VGG16 network with size $h_{i} \times w_{i} \times d_{i}$.

The value for $\lambda$ was chosen to be 0.1 per the recommendation in \cite{8340157}.

\subsection{MSE Based Difference Image Prediction}
Wu et al. simplify the denoising problem by predicting the difference between LDCT and NDCT images instead of reconstructing the NDCT image \cite{Wu2017ACC}. This allows networks to converge much more quickly, which is especially important for conserving training time when cascading multiple networks together. In the first level of the cascade, the inputs are LDCT and the labels are difference images $\mathbf{X}_{L}-\mathbf{X}_{H}$ where $\mathbf{X}_{L}$ is the LDCT image and $\mathbf{X}_{H}$ is the NDCT image. Next, predicted difference images are translated into NDCT predictions $\mathbf{x}_{D}^{(1)}$ by subtracting predicted difference images from LDCT in the first level, and subtracting predicted difference images from NDCT predictions in the following levels. After the first level, Wu et al. concatenate $\mathbf{X}_{L}$ with predictions $\mathbf{x}_{D}^{(i)}$ from the previous level $i$ to be used as input for the next level. In levels  following the first they use $\mathbf{x}_{D}^{(i)}-\mathbf{x}_{H}$ as labels. 

Figure \ref{arch} gives a breakdown of the proposed architecture. First, the DRL network (Figure \ref{model}) is trained on image patches to predict NDCT from LDCT by minimizing $L_{p}(\theta)$. Next, we perform prediction on all image patches and obtain $\hat{Y}$. In order to train the next network, we calculate $\hat{Y}-Y$ to be used as labels and for the inputs we use $\hat{Y}$. 

During testing, we subtract $\hat{Y}$ from $\hat{Y}-Y$ to arrive at the final model prediction $\hat{Y}_{\text {final}}$. 
 
\begin{figure}[htbp]
\centerline{\includegraphics[width=\columnwidth]{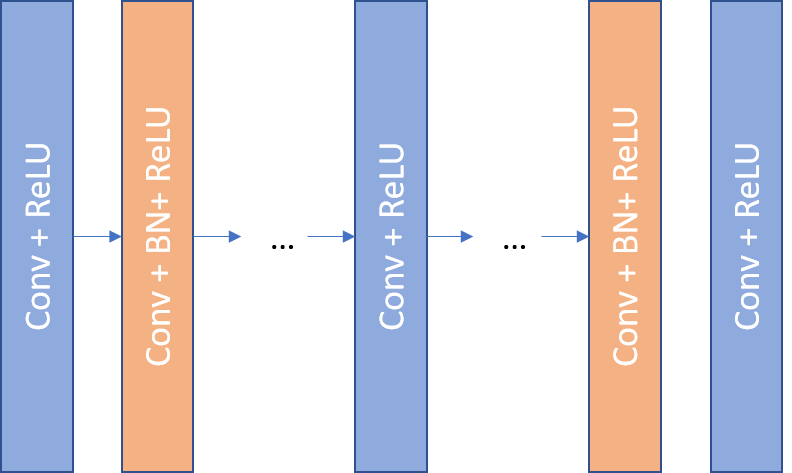}}
\caption{CNN Model Architecture Building Block}
\label{cnn}
\end{figure}

\begin{figure}[htbp]
\centerline{\includegraphics[width=\columnwidth]{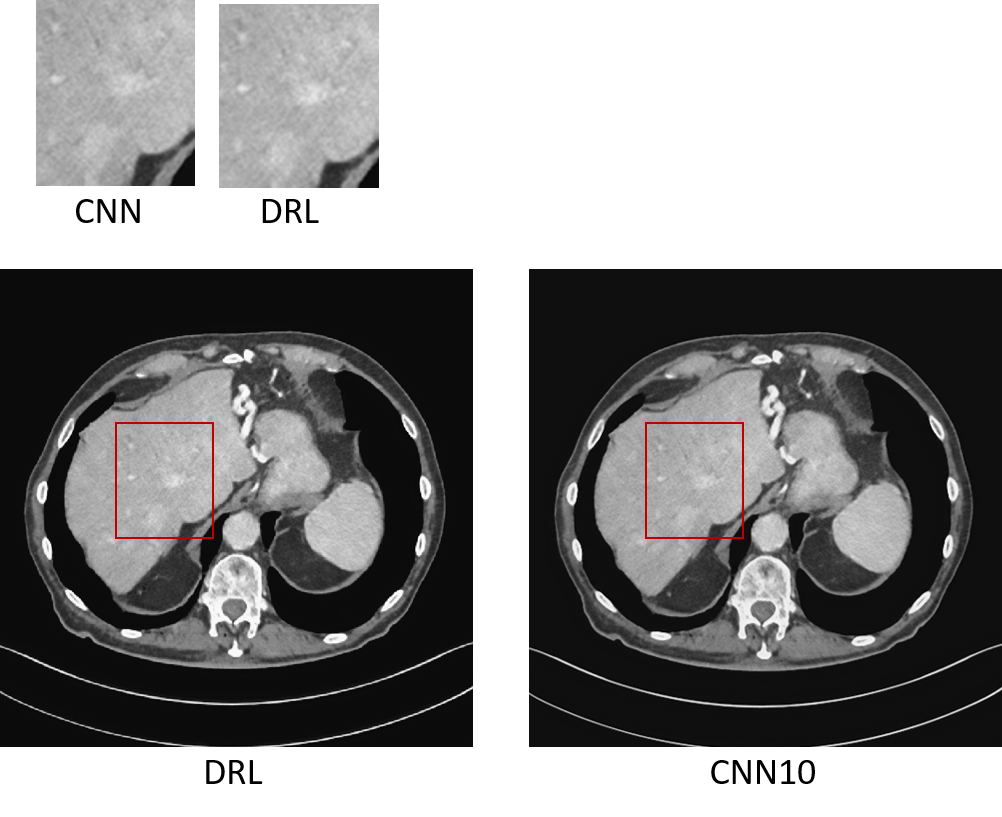}}
\caption{Comparing DRL vs CNN10 Architecture}
\label{drlvscnn}
\end{figure}
 
 \begin{figure}[htbp]
\centerline{\includegraphics[width=\columnwidth]{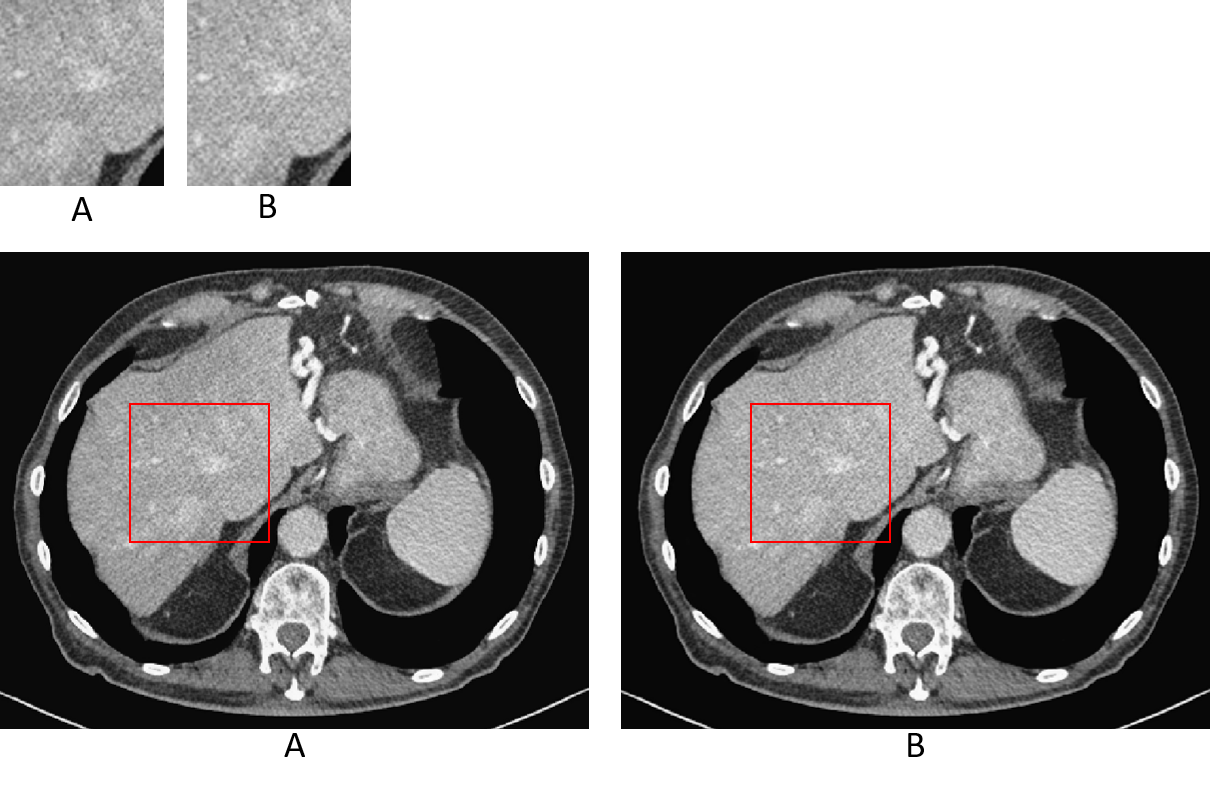}}
\caption{Blended Outputs. A) 2-level cascade B) 3-level cascade }
\label{blend}
\end{figure}
 
 \subsection{Training Details}
 The proposed algorithm is implemented in TensorFlow Keras. The perceptual loss based network is trained for 15 epochs using the Adam optimizer with learning rate 0.001, and the difference image prediction network is trained in the same fashion for 10 epochs. Training is done on a NVIDIA GTX 1070 GPU with a batch size of 32. 
 
\section{Experiments and Results}
First, we compare the basic CNN architecture used in \cite{Wu2017ACC} to the DRL architecture used in \cite{GholizadehAnsari2019DeepLF} to confirm that DRL outperforms when used to predict the difference image. We test it against the CNN10 architecture used in \cite{Wu2017ACC} which consists of 10 Conv + BN + ReLU layers as shown in Figure \ref{cnn} and compare the results to the DRL architecture shown in Figure \ref{model}. The results displayed in Figure \ref{drlvscnn} show a slight improvement in contrast when using the DRL method. The DRL method also enjoys lower parameter count of 225,601 compared to 335, 873 for CNN10. This helps to reduce training and testing time, as well as graphics memory requirements. Based on this finding, we accept the DRL architecture as more effective than a traditional CNN and use it for further experiments. 

In Figure \ref{comp} we show that cascading the difference image predicting technique (A) with the perceptual loss based network (B) with results shown in (C) has better contrast compared to cascading two difference image denoising networks (D) as proposed by \cite{Wu2017ACC}.

Finally, as suggest by Wu et al. we can cascade additional difference image denoising networks to further improve the output. Additionally, Wu et al. generated blended images by calculating a weighted average using the input LDCT image and the predicted image using 0.3 and 0.7 weights respectively. This re-introduces the high frequency texture inherent to CT images that is removed throughout the denoising process. This result is shown in Figure \ref{blend}. Figure \ref{blend}B shows the result of adding an additional cascade to A. Analysing this result shows that our two-level cascade architecture has already reached near optimal denoising because the 3-level cascade result shows negligible improvement. This is preferred in contrast to the architecture used in \cite{Wu2017ACC} which requires many additional cascades for comparable performance. In addition, individual networks in our two-network cascade are less complex than their counter part in \cite{Wu2017ACC} (CNN10).

\section{Discussion and Conclusion}
In this paper we show that cascading a perceptual loss network with a MSE based difference image denoiser enhances contrast and produces images which look closest to the NDCT gold standard when compared to individual networks or cascades without perceptual loss. It is evident from this research, as well as previous works such as \cite{8340157} and \cite{GholizadehAnsari2019DeepLF}, that perceptual loss greatly enhances the ability of the model to preserve structural details during denoising. In conclusion, we have proposed an architecture which outperforms the related works while being less computationally expensive in testing and training. 

\section*{Acknowledgement}
The authors would like to thank Dr. Cynthia McCollough, the Mayo Clinic, the American Association of Physicists in Medicine for making the CT data available for the study. This research has been supported by Mitacs Accelerate grant awarded to Dr. Alirezaie. Authors also gratefully acknowledge the financial support of the Government of Canada.

\bibliography{bibliography.bib}{}
\bibliographystyle{unsrt}

\end{document}